\documentclass[showpacs,showkeys,amsmath,amssymb,useAms]{revtex4}

\usepackage{graphicx}

\def\gs{\mathrel{\lower0.6ex\hbox{$\buildrel {\textstyle >}\over{\scriptstyle \sim}$}}}
\def\ls{\mathrel{\lower0.6ex\hbox{$\buildrel {\textstyle <}\over{\scriptstyle \sim}$}}}

\begin{document}

\title{Limits on decaying dark energy density models from the CMB temperature-redshift relation}

\author{Philippe Jetzer $^{1,2}$}
\email{jetzer@physik.uzh.ch}

\author{Denis Puy $^2$}
\email{puy@graal.univ-montp2.fr}

\author{Monique Signore $^3$}
\email{monique.signore@obspm.fr}

\author{Crescenzo Tortora $^{1}$}
\email{ctortora@physik.uzh.ch}

\affiliation{(1) Institut f\"{u}r Theoretische Physik,
Universit\"{a}t Z\"{u}rich,  Winterthurerstrasse 190, CH-8057
Z\"{u}rich, Switzerland} \affiliation{(2) Groupe de Recherche
d'Astronomie et d'Astrophysique du Languedoc, Universit\'e des
Sciences Montpellier II, GRAAL CC72, F-34095 Montpellier cedex 09,
France} \affiliation{(3) Observatoire de Paris, LERMA, 61 av. de
l'Observatoire, F-75014 Paris, France}


\begin{abstract}
The nature of the dark energy is still a mystery and several models
have been proposed to explain it.
Here we consider a phenomenological model for dark energy decay
into photons and particles as proposed
by Lima \cite{lima1}. He studied the thermodynamic aspects of decaying dark
energy models in particular in the case of a continuous photon creation and/or disruption.
Following his approach, we derive a temperature
redshift relation for the CMB which depends on the effective
equation of state $w_{eff}$ and on the ``adiabatic index'' $\gamma$.
Comparing our relation with the data on the CMB temperature as a function
of the redshift obtained from Sunyaev-Zel'dovich observations and at
higher redshift from quasar absorption line spectra, we find
$w_{eff}=-0.97 \pm 0.034$, adopting for the adiabatic index
$\gamma=4/3$, in good agreement with current estimates
and still compatible with $w_{eff}=-1$, implying that the dark energy content being
constant in time.

\end{abstract}



\maketitle

\section{Introduction}

After the discovery that the cosmic expansion is accelerating \cite{perl, reiss}
and the first cosmic microwave background radiation (CMB) observations of a flat
universe \cite{debernardis}, the current standard model of
cosmology implies the existence of dark energy which accounts for
about 70\% of the total energetic content of the universe,
which according to the observations is spatially flat \cite{spe07}.
The nature of the dark energy is still a mystery (see for instance the review of
\cite{caldwell}).


Several models have been proposed to explain dark energy
\cite{PR03,Pad03,D+05,CTTC06,Caldwell02,PR88,RP88,SS00, ma}. 
An alternative consists to consider a phenomenological decaying
dark energy density with continuous creation of photons
\cite{lima2,puy} or matter \cite{ma}. The dark energy might decay
slowly in the course of the cosmic evolution and thus provide the
source term for matter and radiation. Different such models have
been discussed and strong constraints come from very accurate
measurements of the CMB.

CMB is the best evidence for an expanding Universe starting from
an initial high density state. Within the
Friedmann-Robertson-Walker (FRW) models of the Universe the
radiation after decoupling expands adiabatically and scales as
$(1+z)$, $z$ being the redshift. Depending on the decay mechanism
of the dark energy the created photons could lead to distortions
in the Planck spectrum of the CMB. If the dark energy is
considered as a second fluid component transferring energy
continuously to the material component, the second law of
thermodynamics constrains the whole process in such a way that the
temperature law, that is how the temperature of the CMB changes
with time, can be determined. Such a model has been discussed in
detail by Lima \cite{lima1}, who also established under which
conditions the equilibrium relations are preserved. He also
introduced the concept of an ``adiabatic'' vacuum decay. In a
further paper Lima et al. \cite{lima2} following the approach
outlined in \cite{lima1} derived a temperature redshift
relation law $T=(1+z)^{1-\beta}$ assuming a nonvanishing
source term in the balance equation for particle number, which
depends on a function $\beta$ assumed to be constant. This
parameter can then be determined by fitting the temperature
redshift relation to the data. However, being a purely
phenomenological parameter it cannot be related to more
fundamental quantities such as for instance the effective
equation of state of the dark fluid $w_{eff}$.

In this paper we follow the same approach by Lima
\cite{lima1}, we do not restrict to the case of a universe
dominated by a radiation fluid as in \cite{lima2}, since we leave
the adiabatic index $\gamma$ (which appears in the equation of
state of the fluid including both matter and radiation) free to
vary and in particular we do not adopt their simplifying
assumption of a constant $\beta$ and thus we derive a temperature redshift relation for
CMB which depends on more fundamental quantities such as $w_{eff}$
and $\gamma$.
%
%
We compare our temperature redshift relation with the available
data on the CMB temperature as a function of the redshift obtained
from Sunyaev-Zel'dovich observations and at higher redshift from
quasar absorption line spectra.

The paper is organized as follows. In \S\ \ref{sec:theory} we will
introduce the theoretical background for our model, deriving the
main quantities of interest. In \S\ \ref{sec:data}  the
multi-redshift measurements of the CMB temperature are presented,
while \S \ref{sec:discussion} and \S \ref{sec:conclusions} are
devoted to the discussion of the results, conclusions and future
prospects.

\section{Temperature-redshift relation}\label{sec:theory}

We assume the usual Friedmann-Robertson-Walker (FRW) cosmology and
that the material medium of the Universe contains three different
components: a fluid, which includes radiation and matter (both
baryonic and dark matter) as particular cases, for which we have
the law

\begin{equation}
p = (\gamma - 1) \rho \label{2}
\end{equation}
with $\gamma$ in the interval [0 - 2]. Further we assume a dark
energy, quintessence-like $x$ component, with pressure $p_x$ and
density $\rho_x$ and a 'bare' cosmological constant $\Lambda_0$.
With these components we get for the Einstein field equations

\begin{equation}
8\pi G(\rho+\rho_x)+ \Lambda_0= 3\frac{\dot R^2}{R^2}+3
\frac{k}{R^2}~, \label{11}
\end{equation}

\begin{equation}
8\pi G(p+p_x)- \Lambda_0= -2\frac{\ddot R}{R}- \frac{\dot
R^2}{R^2}-\frac{k}{R^2}~, \label{12}
\end{equation}
where a dot means time derivative. For the dark energy we
assume the relation $p_x=-\rho_x$ (thus setting $w_x=-1$) to hold.
This is for instance the case for
quintessence models in the limit where the scalar field does not depend
on time and thus its time derivative vanishes. In the later stages of the Universe
the time dependence is possibly very weak so that $w_x=-1$ holds
up to small corrections, which we will neglect in the following, thus simplifying
our calculations. This can be viewed as a first order approximation.

Furthermore, we assume that there is no curvature, thus $k=0$, and
take the sum of eqs. (\ref{11}) and (\ref{12}). This way we get

\begin{equation}
8\pi G(\rho+p)=2\frac{\dot R^2}{R^2}-2 \frac{\ddot R}{R} = -2\dot
H~. \label{13}
\end{equation}

Following the paper by Lima et al. \cite{lima2} the energy
conservation equation can be written as

\begin{equation}
\dot\rho +3(\rho+p)H=C_x ~,\label{9}
\end{equation}
where $H=\dot R/R$ is the Hubble parameter and

\begin{equation}
C_x= -\dot\rho_x -3(\rho_x+p_x)H~,\label{91}
\end{equation}
is a term which depends on the dark fluid and acts as a
source term for the $\gamma$-fluid energy. Evidently, if no
interaction between the different fluids exists, then $C_x$ is
null and the standard picture is recovered. $C_x$ can describe
different physical situations such as for instance a thermogravitational quantum
creation theory \cite{LA99} or a quintessence scalar field cosmology
\cite{RP88}.

Assuming as mentioned above the relation
$p_x=-\rho_x$ and writing $\rho_x=\Lambda(t)/(8\pi G)$, we have
\begin{equation}
C_x= -\frac{\dot\Lambda(t)}{8\pi G}~. \label{92}
\end{equation}
On the other hand the equation for the particle number density is
given by

\begin{equation}
\dot n +3nH =\psi ~, \label{7}
\end{equation}
where $n$ is the particle number density and $\psi$ is the
particle source term. Using Gibbs law and well-known thermodynamic
identities, following the derivation given in the paper by Lima et
al. \cite{lima2}, one gets
\begin{equation}
\frac{\dot T}{T} = \left(\frac{\partial p}{\partial \rho}\right)_n
\frac{\dot n}{n} - \frac{\psi}{n T \left(\frac{\partial
\rho}{\partial T}\right)_n} \left[p+\rho- \frac{n
C_x}{\psi}\right]~. \label{1}
\end{equation}

To get a black-body spectrum the second term in brackets in
eq.(\ref{1}) has to vanish, thus

\begin{equation}
C_x = \frac{\psi}{n} \left[ p+\rho \right]~. \label{3}
\end{equation}

Note that $C_{x}$ is null if $p=-\rho$, (this is the
trivial case when only a dark fluid exists) or if $\psi = 0$, which
means that both the dark and normal fluids are separately
conserved. This way we get with eq.(\ref{2})

\begin{equation}
C_x = \frac{\psi}{n}\left[ (\gamma - 1)\rho + \rho
\right]=\frac{\psi \gamma \rho}{n} \label{4}
\end{equation}

and eq.(\ref{1}) becomes

\begin{equation}
\frac{\dot T}{T} = \left(\frac{\partial p}{\partial \rho}\right)_n
\frac{\dot n}{n}~. \label{5}
\end{equation}

With $\left(\frac{\partial p}{\partial \rho}\right)_n =(\gamma
-1)$ one obtains

\begin{equation}
\frac{\dot T}{T} = (\gamma -1) \frac{\dot n}{n}~. \label{6}
\end{equation}

Using the equation for the particle number conservation eq.(\ref{7}) into eq.(\ref{6}) leads to

\begin{equation}
\frac{\dot T}{T} = (\gamma -1) \left[\frac{\psi}{n}-3H\right]~.
\label{8}
\end{equation}

With eqs.(\ref{3}) and (\ref{92}) we get

\begin{equation}
\frac{\dot T}{T} = (\gamma -1) \left[-\frac{\dot\Lambda}{8\pi
G(p+\rho)}-3H\right]~. \label{10}
\end{equation}

We insert (\ref{13}) in eq. (\ref{10}) and obtain

\begin{equation}
\frac{\dot T}{T} = (\gamma -1) \left[\frac{\dot\Lambda}{2\dot
H}-3H\right]~. \label{14}
\end{equation}

As next we integrate eq.(\ref{14})

\begin{equation}
\int^{t_0}_{t_1} \frac{\dot T}{T} dt = (\gamma - 1)
\int^{t_0}_{t_1}\left[\frac{\dot\Lambda}{2\dot H}-3H\right] dt~,
\label{15}
\end{equation}
where $t_0$ denotes the present time and $t_1$ some far instant in
the past. With $H=\dot R/R$ the second term in the bracket on the
right and the term on left hand side can be immediately
integrated. Indeed, if $\dot\Lambda$ vanishes and $\gamma=4/3$ one
gets the usual dependence $T(t)=\frac{R(t_1) T(t_1)}{R(t)}$ for a
radiation fluid. Whereas to carry out the integration of the first
term on the right hand side it is useful to perform a change of
variable from $t$ to $z$ and accordingly
$\frac{dt}{dz}=\frac{-1}{H(1+z)}$. This way we get (with $z_1$
corresponding to the time $t_1$ and $z_0=0$ corresponding to $t_0$
present time)

\begin{equation}
ln \frac{T(z=0)}{T(z_1)}+3(\gamma-1)ln\frac{R(z=0)}{R(z_1)}=\frac{(\gamma-1)}{2} \int_0^{z_1}
\frac{\Lambda^{\prime}}{H^{\prime} H(1+z)} dz~, \label{16}
\end{equation}
where $^{\prime}$ denotes derivative with respect to $z$. We now
assume a power law model for the $\Lambda$ term, thus $\Lambda
= B (R/R_0)^{-m}$ or $\Lambda = B (1+z)^m$, where $B$ is a
constant, which is $B=3 H_0^2(1-\Omega_{m0})$ \cite{ma}, if
the ``bare'' cosmological constant $\Lambda_0$ vanishes.

For such a model the Hubble parameter as a function of $m$ can be
computed \cite{ma} and leads to

\begin{equation}
H(z)= H_0 \left[\frac{(3\Omega_{m0}-m)}{3-m}(1+z)^3 +
\frac{3(1-\Omega_{m0})}{3-m}(1+z)^m \right]^{1/2}~, \label{ma}
\end{equation}
where $\Omega_{m0}$ is related to the present value of the total
matter density: $\rho_{m0}=3 H_0^2 M_{Pl}^2 \Omega_{m0}$,
with $M_{Pl} = (8\pi G)^{-1/2}$. Due to the negligible
energy density accounted for by the radiation, this term has been 
neglected here. As next we insert $H(z)$ and its derivative as taken
from eq.(\ref{ma}) into eq.(\ref{16}) and integrate it, to get
(setting $z_1=z$)

\begin{equation}
T(z) = T_0 \left(\frac{R_0}{R(z)}\right)^{3(\gamma-1)}
exp\left(\frac{B(1-\gamma)}{3H_0^2(\Omega_{m0}-1)}A\right)~,
\label{ma1}
\end{equation}
where
\begin{equation}
A=\left[ln((m-3\Omega_{m0})+m(1+z)^{m-3}(\Omega_{m0}-1))-ln((m-3)\Omega_{m0})\right]~.\label{ma2}
\end{equation}

We can also write eq.(\ref{ma1}) as
\begin{equation}
T(z) = T_0 (1+z)^{3(\gamma-1)}
\left(\frac{(m-3\Omega_{m0})+m(1+z)^{m-3}(\Omega_{m0}-1)}{(m-3)\Omega_{m0}}\right)^{\gamma-1}~.
\label{ma3}
\end{equation}
We inserted in the exponent of eq.(\ref{ma1}) the explicit form of
$B$, thus getting as exponent in the above eq. $(\gamma-1)$.
Notice that for $z=0$: $T(0)=T_0$, whereas for $m=0$ the
expression in the parenthesis is equal to 1 and thus
$T(z)=T_0(1+z)^{3(\gamma-1)}$, which for the canonical value of
$\gamma=4/3$ reduces to the standard expression. Also for $m=3$,
although not a realistic value, the expression in the parenthesis
is non-singular and equal to: $1+\frac{3
ln(1+z)(\Omega_{m0}-1)}{\Omega_{m0}}$.

Since from the analysis by Ma \cite{ma} we
expect $m$ to be small we give here an expression for $T(z)$,
where we keep only the leading linear terms in the expansion in $m$
and neglect all terms of higher order. This way we get from eq.
(\ref{ma3})

\begin{equation}
T(z) = T_0 (1+z)^{3(\gamma-1)}
\left(1+\frac{m}{3}-\frac{m}{3\Omega_{m0}}(1+(1+z)^{-3}(\Omega_{m0}-1))\right)^{(\gamma-1)}~.
\label{d2}
\end{equation}

If $m$ is positive, then the
dark energy slowly decreases as a function of the
cosmic time, whereas if $m$ is negative the inverse process happens.

It is easy to calculate the equivalent effective dark energy
equation of state $p=w_{eff}\rho$ with $w_{eff}=\frac{m}{3}-1$.
If $m>0$ then we have $w_{eff} > -1$, i.e. our model is
quintessence-like \cite{PR88,RP88,D+05}, while we have a
phantom-like \cite{Caldwell02} model when $m$ is negative and
$w_{eff}<-1$. Another interesting quantity is the
deceleration parameter, which can be written as
\begin{equation}
q(z)= - \frac{\ddot{R}R}{\dot{R}^{2}}=
\frac{(1+z)^{3}(m-3\Omega_{m0})+3(m-2)(1+z)^{m}(\Omega_{m0}-1)}{2(1+z)^{3}(m-3\Omega_{m0})+6(1+z)^{m}(\Omega_{m0}-1)}~.
\end{equation}
Imposing that $q(z)=0$, we can determine the {\it transition
redshift}, i.e. the redshift when the Universe was changing from a
deceleration to an acceleration phase, which is given by
\begin{equation}
z_{T}=\bigg ( \frac{3(2-m)(1-\Omega_{m0})}{3\Omega_{m0}-m}
\bigg)^{\frac{1}{3-m}}-1~.
\end{equation}
From this result we have that the larger $m$ is, the earlier the
Universe changes from deceleration to acceleration.



\section{Multi-redshift measurements of $T_{\rm CMB}$ and fitting
procedure}\label{sec:data}

To test the decaying $\Lambda$ model we will rely on the CMB
temperatures derived from the absorption lines of high redshift
systems and the ones from Sunyaev-Zel'dovich effect in clusters of
galaxies (we will collectively quote as $T_{CMB}$,
hereafter).

%

We can infer the CMB temperature at high redshift from the
analysis of quasar absorption line spectra which give atomic or
ionic fine structure levels excited by the photo-absorption of the
cosmic microwave background radiation. Detection of absorption
from the ground states and excited of C${\rm_I}$ in damped
Ly$\alpha$ system towards quasars permits to estimate the
population ratio of the excited fine-structure levels, and thus to
derive the CMB temperature at the redshift of the damped
Ly$\alpha$ system. Ge et al. \cite{ge97}, in the damped Ly$\alpha$
system of the QSO 0013-004, found the CMB temperature

\begin{equation}
T_{\rm CMB}=7.9 \pm 1.0 \ {\rm K} \ {\rm at} \ z=1.9731.
\end{equation}

Srianand et al. \cite{sri00} detected absorption lines in an
isolated gas cloud toward PKS1232+0815, with which they could get
the following limits on the CMB temperature

\begin{equation}
6.0<T_{\rm CMB} < 14 \ {\rm K} \ {\rm at} \  z=2.33771,
\end{equation}
and Molaro et al. \cite{mol02} in the damped Ly$\alpha$ system
toward QSO 0347-3819 gave

\begin{equation}
T_{\rm CMB} = 12.1^{+1.7}_{-3.2} \ {\rm K} \ {\rm at} \  z=3.025.
\end{equation}

The cosmic microwave background can also excite levels of
molecular species, when the energy separation involved corresponds
to the CMB peak frequency. We know that molecular gas is an
important ingredient of star formation. Damped Ly$\alpha$
absorbers (i.e. DLAs) are generally taken as seeds of present-day
galaxies, and for this reason DLAs can be considered as an
important gas reservoir for star formation. H$_2$, the most
abundant molecules in the Universe is a good candidate (see Puy et
al. \cite{puy93}, Galli \& Palla \cite{gal98} and Stancil et al.
\cite{sta98}). The search of H$_2$ in DLAs can be carried out by
observing the H$_2$ absorption lines in the lyman
X$^1$$\Sigma^+_{\rm g} \longmapsto$ B$^1$ $\Sigma^+_{\rm u}$. Cui
et al. \cite{cui05} obtained the spectrum of molecular hydrogen
associated with the damped Ly$\alpha$ system at $z=1.7765$ toward
the quasar Q1331+170 and constructed a model to describe the
structure of H$_2$ absorber. Applying the inferred conditions to
the C$_{\rm I}$ fine structure excitation, they found the CMB
temperature to be
\begin{equation}
T_{\rm CMB} = 7.2 \pm 0.8 \ {\rm K} \ {\rm at} \  z=1.7765.
\end{equation}
Srianand et al. \cite{sri08}, from the CO rotational excitation
temperatures, derive in a damped Ly$\alpha$ system towards SDSS
J143912.04+111740.5
\begin{equation}
T_{\rm CMB} = 9.15 \pm 0.72 \ {\rm K} \ {\rm at} \  z=2.41837.
\end{equation}

During passage through a cluster of galaxies some of the photons
of the cosmic microwave background radiation are scattered by
electrons in the hot intracluster medium. This imprint was first
described by Sunyaev-Zel'dovich \cite{sun72}. Thus, spectral
measurements of galaxy clusters at different frequency bands yield
independent intensity ratios for each cluster. The combinations of
these measured ratios permit to extract the cosmic microwave
background radiation (see Fabbri et al. \cite{fab78}). Recently,
Luzzi et al. \cite{luz09} have analyzed the results of
multifrequency Sunyaev-Zel'dovich measurements toward several
clusters from 5 telescopes (BIMA, OVRO, SUZI II, SCUBA and MITO),
see Table 1.

\begin{table}
\caption{CMB $T_{\rm CMB}$
values for several clusters from Luzzi et al. \cite{luz09}.}
\begin{minipage}[htbp]{\columnwidth}
\begin{tabular}{lccc}
\hline \hline Cluster & $T_{\rm CMB}$ & $\ z$ \\
&  (K)\\ \hline

     A1656 & 2.72 $\pm$  0.10 &   $\ $ 0.023 \\

     A2204 & 2.90 $\pm$  0.17 &   $\ $ 0.152 \\

     A1689 & 2.95 $\pm$  0.27 &   $\ $ 0.183 \\

      A520 & 2.74 $\pm$  0.28 &   $\ $ 0.200 \\

     A2163 & 3.36 $\pm$  0.20 &   $\ $ 0.202 \\

      A773 & 3.85 $\pm$  0.64 &   $\ $ 0.216 \\

     A2390 & 3.51 $\pm$  0.25 &   $\ $ 0.232 \\

     A1835 & 3.39 $\pm$  0.26 &   $\ $ 0.252 \\

      A697 & 3.22 $\pm$  0.26 &   $\ $ 0.282  \\

    ZW3146 & 4.05 $\pm$  0.66 &   $\ $ 0.291 \\

   RXJ1347 & 3.97 $\pm$  0.19 &   $\ $ 0.451 \\

 CL0016+16 & 3.69 $\pm$  0.37 &   $\ $ 0.546 \\

    MS0451 & 4.59 $\pm$  0.36 &   $\ $ 0.550 \\

\hline

\end{tabular}
\end{minipage}
\end{table}


We will match the observed $T_{CMB}$ with the theoretical
expression $T_{th}$, which we have derived in eq.
(\ref{ma3}), by minimizing the following merit function
\begin{equation}
\chi^{2}_{TCMB}= \sum_{i=1}^{N_{TCMB}} \bigg
(\frac{T_{th}^{i}- T_{CMB}^{i}}{\sigma_{CMB,i}} \bigg)^{2}~,
\end{equation}
where $\sigma_{CMB,i}$ is the error on the temperature estimates
and $N_{TCMB}=18$ is the number of available observational
data. We will find the best fitted parameters which correspond to
the minimum of $\chi^{2}$, i.e. $\chi^{2}_{min}$ and we will
determine the $68\%$ uncertainties by imposing $\Delta \chi^{2}=
\chi^{2}-\chi^{2}_{min} = 1$.

\section{Results}\label{sec:discussion}

We have tested our model by comparing the CMB temperature
predicted (see eq. \ref{ma3}), with the collection of
multi-redshift measurements of $T_{\rm CMB}$ we have discussed in
the previous section. We set $T_0 \, = \, 2.725 \ {\rm K}$,
which is quite well determined in the literature \cite{mat99},
and the matter density $\Omega_{\rm m0}=0.273$
to the value inferred in Komatsu et al. \cite{komatsu+09}.

\begin{figure*}[h!tbp]
\centering
\includegraphics[scale=0.7,angle=0]{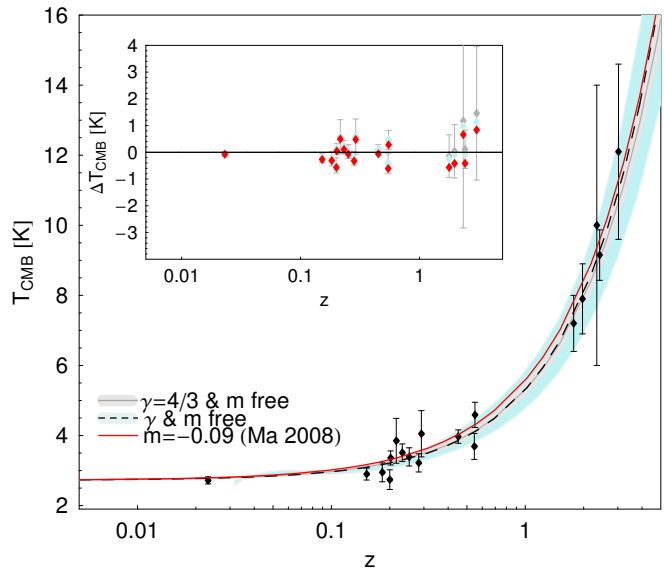}
\caption{Cosmic microwave background temperature for different
m-models. The data are plotted as black symbols. As shown in the
legend, the gray line and the lighter gray region are the best
fitted curve and $1\sigma$ scatter for the case with $\gamma=4/3$
and $m$ free. The black dashed line with the cyan region are
relative to the case with both $\gamma$ and $m$ free. Finally the
red curve is the best fit curve when using the value for m as
found by Ma \cite{ma}. In the inset, we show the difference
between the data points and each best fitted curve as a function
of redshift.}\label{fig:fig1}
\end{figure*}

If we take $\gamma=4/3$, then we find $m=0.09 \pm 0.10$ and $z_{T}
= 0.82 \pm 0.10$, while, if we leave both $\gamma$ and $m$ free to
vary $m=0.20_{-0.23}^{+0.23}$ and $\gamma = 1.35_{-0.05}^{+0.03}$
and $z_{T}=1.02\pm 0.23$. In both the cases, the best fitted $m$
values are positive, but consistent with $0$ within $1\sigma$
uncertainty, while in the latter case the estimated value for
$\gamma$ of $1.35$ is consistent within the errors with the
canonical value for radiation. Within the uncertainties, the
derived $z_{T}$ are consistent with the typical values in
literature (e.g. \cite{CTTC06,Daly+08}), although, on
average, higher. In Fig. \ref{fig:fig1} we plot the $T-z$ diagram
with the different best fit curves.

Assuming a constant value for the ratio $\Psi/3nH=\beta$, with the
constraint $0 \leq \beta \leq 1$, Lima et al. \cite{lima2} found
as a result instead of our eq.(\ref{ma3}) the relation
\begin{equation}
T(z)=T_0 (1+z)^{1-\beta}~. \label{d1}
\end{equation}
For a value of $\beta$ different from zero the temperature of the
expanding universe at high values of $z$ is slightly lower than in
the standard photon-conserved scenario. Lima et al. \cite{lima2}
discussed the various upper limits which can be derived on $T(z)$.
The best values are provided by absorption lines from molecules.
The results are presented in Fig. 1 of their paper, from which one
can see that the upper limits are well compatible with the
standard photon-conserved scenario, i.e. $\beta=0$. Luzzi et al.
\cite{luz09} found using their data (as given in our Table 1) and
also the values mentioned in eqs. (23)-(26) as best fit the value
$\beta=0.024^{+0.068}_{-0.024}$. We performed, as a check, the
same fit using the same data with in addition the value given in
eq.(27) and got $\beta=0.027^{+0.04}_{-0.027}$, which clearly is
consistent with their value. Fitting several data sets to
constrain the $H(z)$ function (as given by eq. \ref{ma}), Ma
\cite{ma} obtained $m=-0.09^{+0.08}_{-0.11}$ and
$\Omega_{m0}=0.29^{+0.03}_{-0.07}$. This model with negative $m$
is plotted as well in Fig. \ref{fig:fig1} as a comparison,
although it is ruled out by our fitting of the CMB.

\section{Conclusions} \label{sec:conclusions}

We studied a model for the dark energy decay based on the
assumption that the dark energy is regarded as a second fluid
component transferring energy continuously to the material
component as suggested by Lima \cite{lima1}. In which case the
second law of thermodynamics constrains the whole process, thus
allowing to determine the temperature redshift relation. As a new
point we were able to derive for the latter quantity a relation
depending on parameters such as $w_{eff}$ and $\gamma$. Thanks to
recent CMB temperature measurements via the Sunyaev-Zel'dovich
effect and others at higher redshift from quasar absorption lines,
we could determine best fit values for these parameters. In
particular, assuming the canonical 4/3 value for $\gamma$ we found
$m=0.09 \pm 0.10$ corresponding to $w_{eff}=-0.97 \pm 0.034$,
whereas leaving $\gamma$ open we got $m=0.20 \pm 0.10$ or
$w_{eff}=-0.93 \pm 0.08$. Both results are within 1$\sigma$
consistent with the canonical value of $-1$ for the cosmological
constant. Using SNe Ia data combined with CMB and Baryon Acoustic
Oscillations, and assuming a flat universe, Kowalski et al.
\cite{kowalski} found $w_{eff}=-0.97 \pm 0.06 \pm 0.06$ (stat,
sys) and similarly Kessler et al. \cite{kessler} estimated
$w_{eff}=-0.96 \pm 0.06 \pm 0.12$. Our best fit value for
$w_{eff}$ compares quite well with these values, and we point out
that our ``derivation'' is completely different and to our
knowledge new. The values $z_T$ we find for the transition
redshift are in good agreement within the uncertainties with the
typical ones quoted in the literature.

Clearly, it would be nice to have better measurements for the CMB
temperature at high redshift, since the difference between the
different models increases for high $z$ values. We notice also
that since the best fit value for $m$ is small, it implies that
the dark energy variation with time is very slow (at least when
not considering too high $z$ values, where we expect our
approximations to break down) and thus our assumption on $w_x=-1$
can be considered as justified a posteriori.


\noindent
Acknowledgements

The authors thank Y. Chastagnier and  
Profs M. De Petris, Y. Rephaeli and N. Straumann for useful
discussions. Ph. Jetzer, thanks the French Regional Council of
Languedoc Roussillon and the Physics Institute of Montpellier (IPM) for the invitation at the
Laboratory of Astrophysics of the University of Montpellier. M. Signore
thanks the Programme National de Cosmologie et Galaxie (PNCG) and the Swiss National Science
Foundation for her many visits at the University of Montpellier
and the University of Zurich. C. Tortora was supported by
the Swiss National Science Foundation.

\end{document}